\documentclass[11pt,twoside]{article}
\usepackage{amsmath}
\usepackage{asp2010}

\resetcounters
\begin{document}

\title{Radiation-MHD simulations of pillars and globules in HII regions}
\author{Jonathan~Mackey$^{1,2}$
\affil{$^1$Argelander-Institut f\"ur Astronomie, Auf dem H\"ugel 71, 53121 Bonn, Germany}
\affil{$^2$Dublin Institute for Advanced Studies, 31 Fitzwilliam Place, Dublin 2, Ireland}}

\begin{abstract}
Implicit and explicit raytracing--photoionisation algorithms have been implemented in the author's radiation-magnetohydrodynamics code.
The algorithms are described briefly and their efficiency and parallel scaling are investigated.
The implicit algorithm is more efficient for calculations where ionisation fronts have very supersonic velocities, and the explicit algorithm is favoured in the opposite limit because of its better parallel scaling.
The implicit method is used to investigate the effects of initially uniform magnetic fields on the formation and evolution of dense pillars and cometary globules at the boundaries of HII regions.
It is shown that for weak and medium field strengths an initially perpendicular field is swept into alignment with the pillar during its dynamical evolution, matching magnetic field observations of the`Pillars of Creation' in M16.
A strong perpendicular magnetic field remains in its initial configuration and also confines the photoevaporation flow into a bar-shaped, dense, ionised ribbon which partially shields the ionisation front.
\end{abstract}

\section{Introduction}
Motivated by observations of ongoing star formation within pillars and globules at the borders of H~\textsc{ii} regions \citep[e.g.][]{HesScoSanEA96}, and by increasing evidence that star formation propagates outwards from young clusters sequentially \citep{LeeChe07,SmiPovWhiEA10}, models for the formation and evolution of theses structures have been investigated in increasing detail recently.
These are highly non-linear structures forming in the presence of intense radiation fields, so multidimensional numerical simulations including hydrodynamics (HD) with photoionisation (R-HD) are required, ideally also including self-gravity and magnetohydrodynamics (MHD).
Low resolution 2D calculations were already performed in the 1980s \citep{Yor86}; ionisation front instabilities were shown to produce structures resembling pillars during the expansion of H~\textsc{ii} regions from the ultra-compact phase \citep{GarSegFra96}, also shown more recently in 3D simulations \citep{WhaNor08}.
Shadowing due to pre-existing overdensities was also shown to generate pillar-like structures in R-HD simulations \citep{WilWarWhi01}; the effectiveness of this process in a turbulent density field was demonstrated by \citet{MelArtHenEA06} and \citet{GriBurNaaEA10}, and in the presence of magnetic fields by \citet{ArtHenMelEA11}.
In \citet{MacLim10} we studied the formation of massive pillars due to shadowing of randomly placed dense clumps with 3D R-HD simulations, showing that this simple model could reproduce the main features of optical images, the morphology of molecular emission, and the line-of-sight velocity profiles of the M\,16 pillars obtained from molecular line emission \citep{Pou98}.
In this contribution simulations performed by the author and Dr.\ Andrew J.\ Lim are described which include MHD with non-equilibrium photoionisation and raytracing (R-MHD), with the aim of determining the effects of magnetic fields on the pillar formation process studied in \citet{MacLim10}.
This work is described in more detail in \citet{MacLim11b} and is complementary to R-MHD simulations of global H~\textsc{ii} region evolution by \citet{ArtHenMelEA11} and photoionisation of single clumps by \citet{HenArtDeCEA09}.
In the following section the R-MHD code used for the calculations is described and its scaling behaviour on parallel supercomputers is discussed for implicit and explicit raytracing--photoionisation algorithms.
In section \ref{sec:results} the R-MHD simulations are described and the results obtained are summarised.

\section{Code Description and Parallel Scaling}
\label{Code}
In our R-MHD code \citep{MacLim10,MacLim11b} the equations of ideal compressible MHD are solved on a uniform grid using a directionally-unsplit, second order (in time and space), finite volume formulation with a Roe-type Riemann solver in conserved variables. 
A tracer variable is used for the ion fraction of H; 
microphysical (radiative and collisional) heating and cooling processes also provide a source term to the energy equation and are included by operator splitting.
The column density from a radiation source to a given grid cell is calculated using a short characteristics raytracing module; diffuse radiation is treated approximately by the on-the-spot approximation.
Photoionisation heating and recombination cooling are calculated explicitly for H using its non-equilibrium ion fraction;
cooling due to other elements is approximate and uses model C2 in \citet{MacLim10}.
The code is parallelised using MPI and was run on 128 cores for the calculations in section \ref{sec:results}.

An implicit raytracing algorithm was used for the simulations presented here, based closely on the C$^2$-ray algorithm of \citet{MelIliAlvEA06} but with some modifications
(referred to here as Algorithm 1 or just Alg1).
This updates microphysical quantities (internal energy, ion fraction) as rays are being traced outwards from the source, using time-averaged attenuation fractions through cells and thus allowing ionisation fronts to cross many optically thick grid cells in a single raytracing step without loss of photon conservation.
The only serious disadvantage of this integration scheme is that it is difficult to parallelise efficiently on non-spherical grids: raytracing outwards from a source must proceed sequentially from one processor's subdomain to its neighbour and so on.
Since the microphysical quantities are integrated during raytracing this places a significant fraction of the total computation in a poorly-scaling algorithm.
In the limit of many subdomains the parallel scaling of this algorithm is simple to deduce: the radial direction must be calculated (at least partially) in serial and only the perpendicular direction(s) can be efficiently parallelised.
In 2D, therefore, the strong scaling of the raytracing algorithm on $N$ processes will have runtime $t_{\mathrm{sim}}\propto N^{-1/2}$, and in 3D the scaling will be $t_{\mathrm{sim}}\propto N^{-2/3}$.
This scaling holds for the short characteristics tracer (and probably also for other ray-splitting algorithms).

\begin{figure}
\centering
\resizebox{0.49\hsize}{!}{\includegraphics{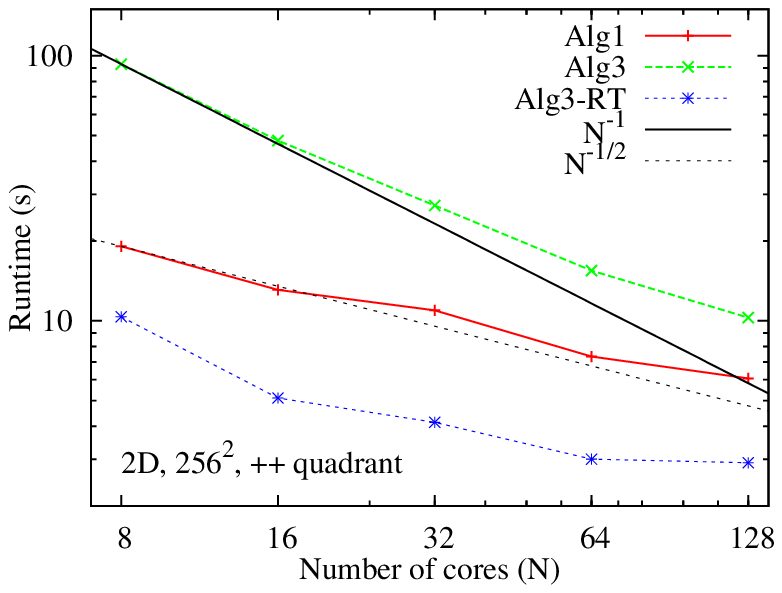}}
\resizebox{0.49\hsize}{!}{\includegraphics{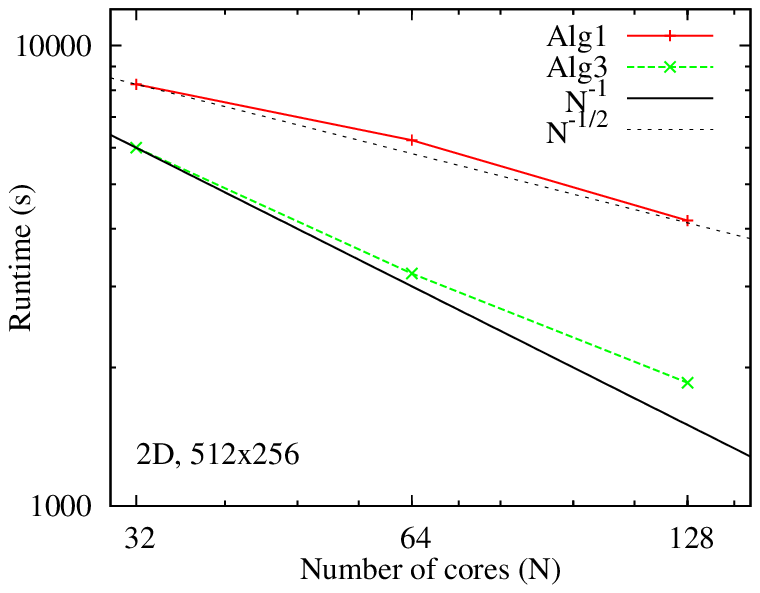}}
\caption{
  Scaling comparison between Algorithms 1 and 3 on the distributed memory cluster JUROPA.
  Left: Scaling of a 2D static problem as discussed in the text.
  Right: Scaling for a 2D calculation with photoionisation and a $2000\,\mathrm{km}\,\mathrm{s}^{-1}$ stellar wind, as discussed in the text.
  Timing was measured with a microsecond timer in the main simulation loop, not including time spent in initialisation or data output.}
\label{fig:scaling}
\end{figure}

An explicit raytracing algorithm has recently been added to the code, similar to the method of~\citet{WhaNor06}, hereafter denoted Alg3.
Here only the column densities are calculated in the raytracing step and the microphysics integration is performed fully in parallel. 
A second order integration, using the same approach as for the hydrodynamics, proved much more efficient than the usual first order method.
A raytracing is performed at the start of the timestep; everything is integrated forwards half a timestep; another raytracing is performed using this time-centred density (and ion fraction) field; these column densities are then used to integrate the full timestep from the initial to time-advanced values.
With this algorithm only 8 raytracings (4 timesteps) are required to photoionise an optically thick cell with an error of less than 1\%~\citep{Mac11}; the microphysics timestep limit used is $\Delta t = 0.25/\dot{x}$, where $\dot{x}$ is the time rate of change of the H$^+$ fraction, $x$.

The total runtime for the code in 2D simulations (cylindrical coordinates $[R,z]$) as a function of number of processors $N$ is shown in Fig.~\ref{fig:scaling} for two different calculations.
The first simulation is of the expansion of an H~\textsc{ii} region into a uniform static medium covering the range $[R,z]\in[0,2]$ pc with a constant gas density of $n_{\mathrm{H}}=100\,\mathrm{cm}^{-3}$ and initial temperature $T=50$K.
The star at the origin has monochromatic ionising photon luminosity $\dot{N}=10^{48}\,\mathrm{s}^{-1}$.
The grid contains $256^2$ cells and the simulation was run for 10 recombination times ($t_{\mathrm{rec}}$, here $3.861\times10^{10}$ s) on $8-128$ processors (Fig.~\ref{fig:scaling}, left panel).
The better scaling properties of the explicit Alg3 are evident, but the overall efficiency of Alg1 is so much greater that it remains the faster algorithm.

In the second simulation the star at the origin has luminosity $\dot{N}=3\times10^{48}\,\mathrm{s}^{-1}$ and a fast stellar wind is injected in a 10 cell radius around the origin with mass-loss rate $\dot{M}=10^{-7}\,\mathrm{M}_{\odot}\,\mathrm{yr}^{-1}$ and velocity $v_{\mathrm{w}}=2000\,\mathrm{km}\,\mathrm{s}^{-1}$.
The model is run with $512\times256$ grid cells and physical domain $z\in[-1.28,1.28]\times10^{18}$ cm, $R\in[0,1.28]\times10^{18}$ cm, with ambient density $n_{\mathrm{H}}=3000\,\mathrm{cm}^{-3}$.
Here the overall timestep is set by the Courant condition on the $2000\,\mathrm{km}\,\mathrm{s}^{-1}$ wind region and not by the microphysics.
The simulation is run for 10 kyr, or $\simeq245t_{\mathrm{rec}}$.
The right panel of Fig.~\ref{fig:scaling} shows that Alg3 is more efficient than Alg1 for this problem on $32-128$ cores.
This new algorithm will be used for calculations of the circumstellar medium around evolving massive stars.

\begin{figure}
\centering
\resizebox{0.75\hsize}{!}{\includegraphics{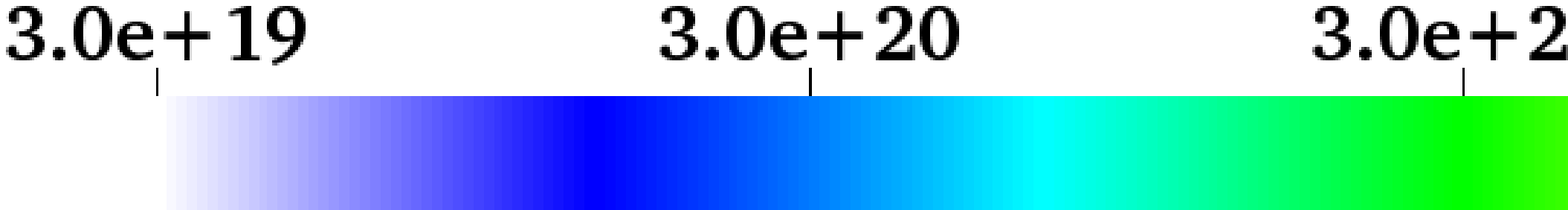}}\\
\resizebox{0.24\hsize}{!}{\includegraphics[angle=-90]{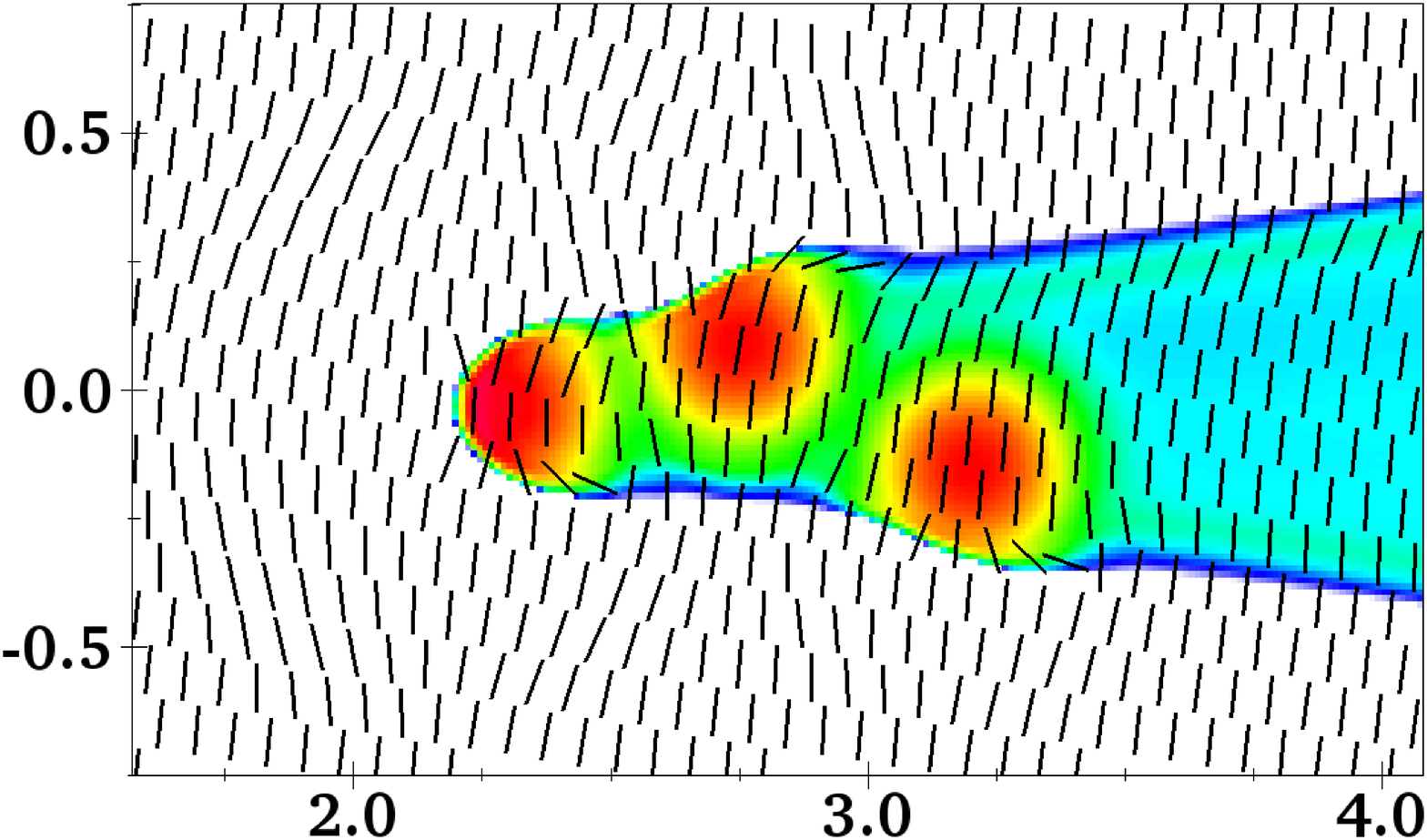}}
\resizebox{0.24\hsize}{!}{\includegraphics[angle=-90]{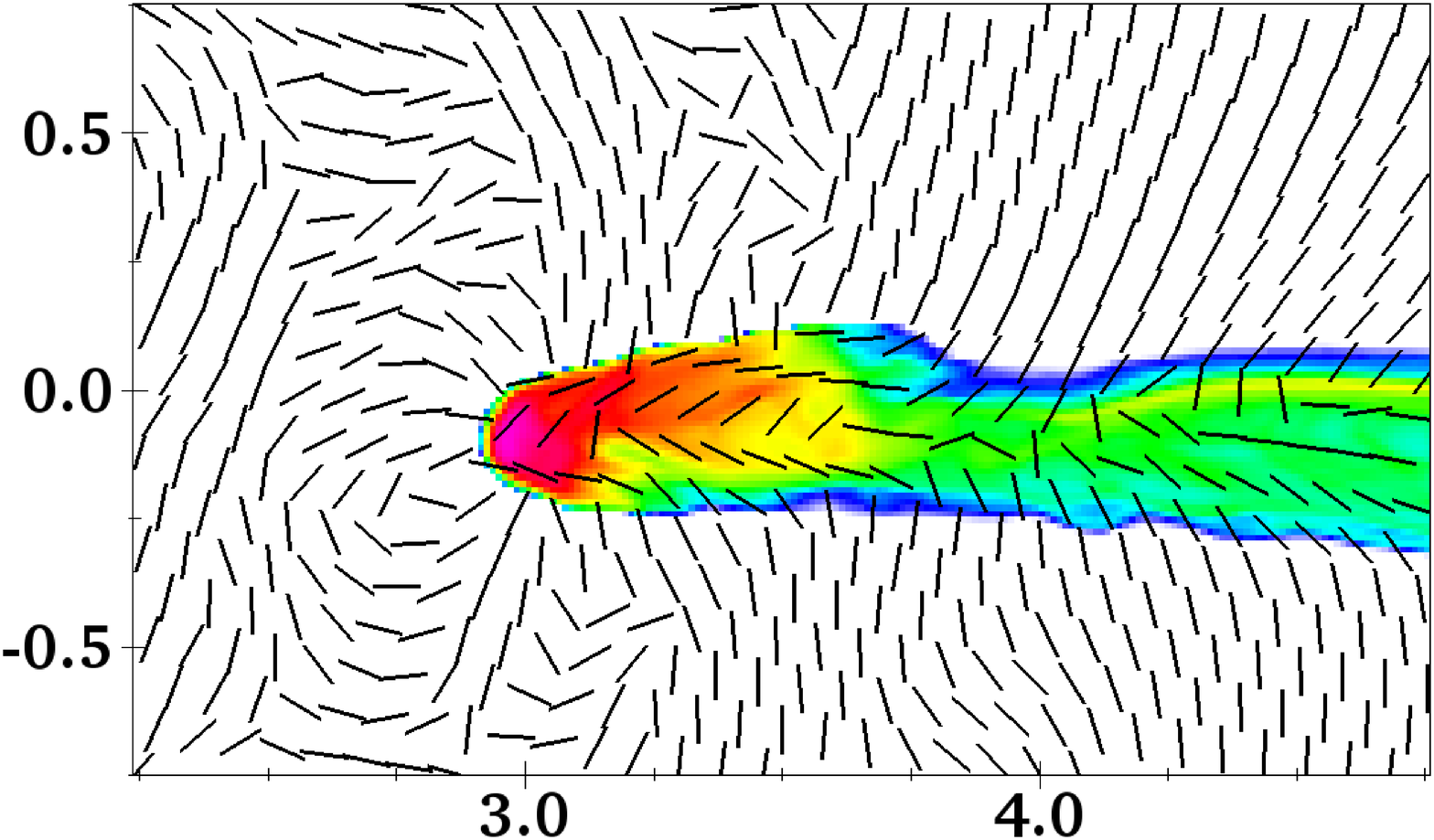}}
\resizebox{0.24\hsize}{!}{\includegraphics[angle=-90]{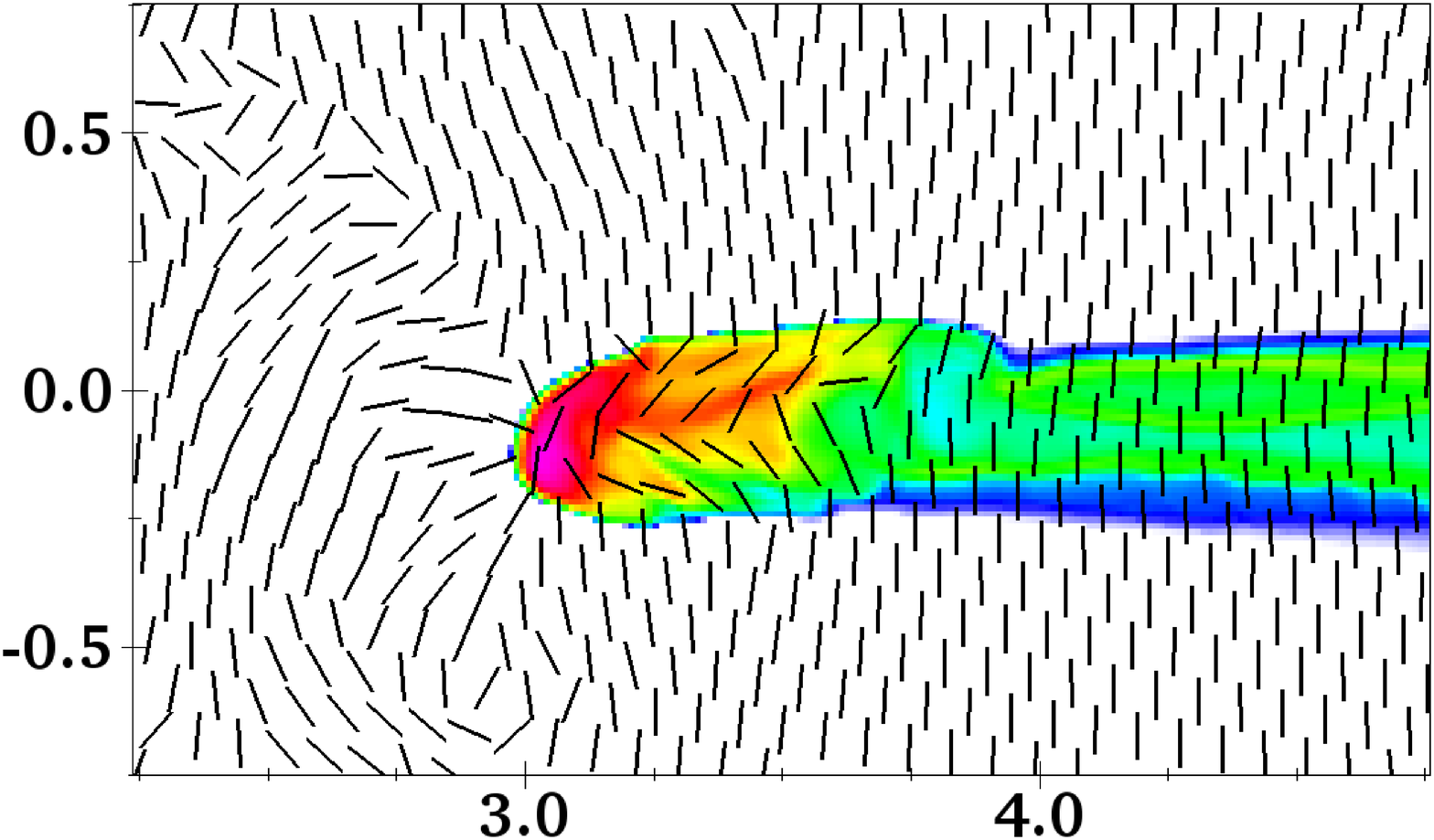}}
\resizebox{0.24\hsize}{!}{\includegraphics[angle=-90]{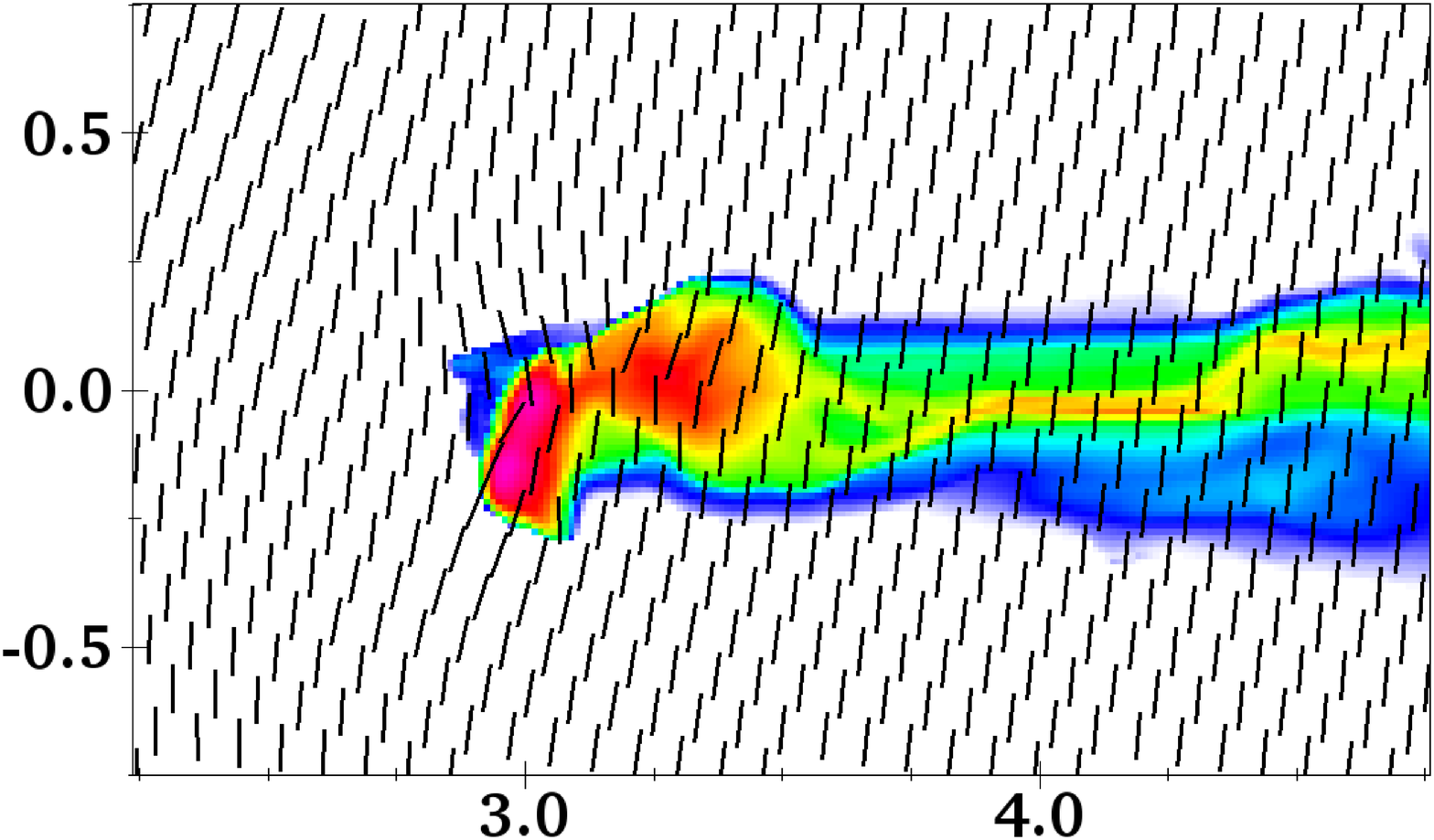}}
\caption{
  Projected gas density ($x$-$z$ plane) for models R2, R5 and R8 on the logarithmic column density scale indicated.
  The radiation source is located above the simulation domain at $[0,0,0]$.
  Projected magnetic field orientation is overlaid (weighted by gas density in the line-of-sight integral, and recovered from Stokes Q and U parameters).
  The left panel shows R2 at $t=25$ kyr, the other three show R2, R5, and R8 (from left to right) at $t=250$ kyr.
  }
\label{fig:results}
\end{figure}

\section{R-MHD Simulation Results}
\label{sec:results}
We have performed a number of 3D R-MHD simulations of the formation and evolution of pillars and globules, presented in more detail in~\citet{MacLim11b}.
Here we present results from simulations involving the photoionisation of three dense clumps of gas embedded in a uniform ambient medium and exposed to ionising radiation from a point source.
The ambient gas density is $n_{\mathrm{H}}=200\,\mathrm{cm}^{-3}$, and the clumps have peak densities $500\times$ greater with Gaussian density profiles [$\rho\propto\exp(-r^2/2r_0^2)$] of width $r_0=0.09$ pc.
The 3D simulation domain is $x\in[1.5,6.0]$ pc, $\{y,z\}\in[-1.5,1.5]$ pc resolved by $384\times256^2$ grid cells; the radiation source is off the grid at the origin with monochromatic photon luminosity $\dot{N}=2\times10^{50}\,\mathrm{s}^{-1}$.
The three clumps are almost collinear with positions (in pc) $[2.3,0,0]$, $[2.75,0,0.12]$ and $[3.2,0,-0.12]$.
Three simulations are presented here: R2 has an initial magnetic field $\mathbf{B}=[1.8,1.8,17.7]\,\mu$G, R5 has $\mathbf{B}=[0,0,53.2]\,\mu$G, and R8 has $\mathbf{B}=[14.2,7.1,159]\,\mu$G.
All three models have a field largely perpendicular to the radiation propagation direction.
The weak $\mathbf{B}$ field in R2 means that the results closely follow purely hydrodynamic evolution (see fig.~7 in \citealt{MacLim10}, describing model 17), whereas R8 is largely magnetically dominated; R5 lies between the two extremes.
All models were evolved to at least 400 kyr.

The projected neutral hydrogen column density is shown in Fig.~\ref{fig:results} at times $t=25$ kyr (for R2) and $250$ kyr (R2, R5, R8), with projected magnetic field orientation indicated by the overlaid black lines.
At 25 kyr all three models are effectively identical -- an R-type ionisation front has ionised most of the low density gas on the grid except for the shadowed region and the three clumps are beginning to be compressed by shocks generated by the high pressure ionised gas.
By 250 kyr, however, clear differences have emerged: the weak field model R2 behaves almost identically to a purely hydrodynamic simulation but the strong field model R8 is very diffeent.
In dense neutral gas in R2 the initially perpendicular magnetic field has been swept into alignment with the long axis of the dense pillar.
The projected $\mathbf{B}$ field in the ionised gas around the head of the pillar follows the radial photoevaporation flow generated at the ionisation front; this is also seen in model R5 but is absent in R8.
The alignment of the magnetic field with the long axis of the pillar is also weaker in R5 and the field orientation is essentially unchanged in R8.
The projected density structure is rather different in R8 also, in that the clump furthest from the source has stayed largely intact and separate from the two closer to the source (which have merged).
In R2 an R5 the furthest clump is no longer distinguishable from the general structure of the pillar.

\begin{figure}
\centering
\resizebox{0.9\hsize}{!}{\includegraphics{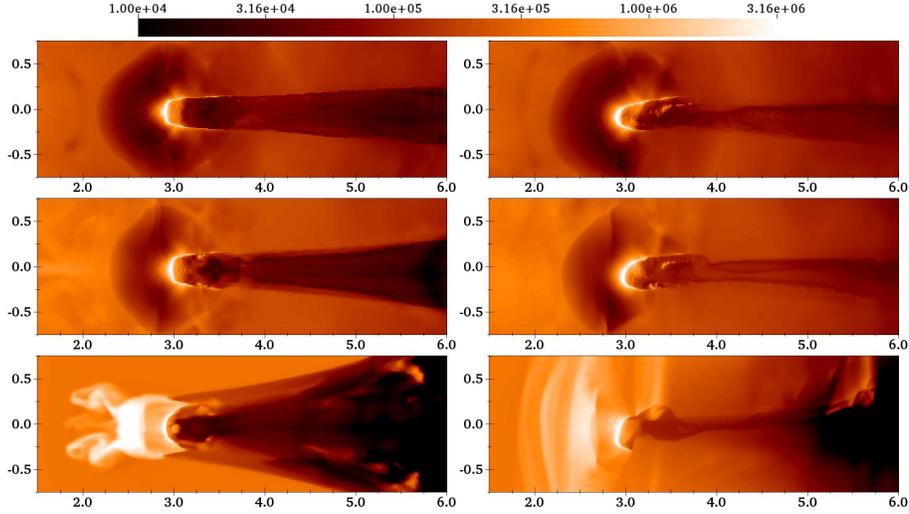}}
\caption{
  Simulations R2 (above), R5 (centre) and R8 (below) at $t=250$ kyr, shown here in H$\alpha$ emission (arbitrary logarithmic intensity scale).
  The left sequence shows projections along the $z$ axis (parallel to initial magnetic field orientation), while the right hand side shows projections along the $y$ axis, perpendicular to the field and the same projection as in Fig.~\ref{fig:results}.
  }
\label{fig:emission}
\end{figure}

R8 is also different in that it has significant neutral low-column-density gas (dark blue).
This can be understood by looking at the ionised gas emission in Fig.~\ref{fig:emission} which shows that in R8 there is a lot of dense ionised gas ahead of the pillar.
This provides significant optical depth and allows gas to recombine in its shadow around the pillar.
This gas has been organised by the strong magnetic field into a bar-shaped or filamentary structure.
The photoevaporation flow is clearly seen in these figures for models R2 and R5, but for R8 it is confined to a much smaller volume, set by the radius at which the ram pressure of the flow equals the magnetic pressure in the ionised gas, at which point an isothermal reverse shock forms.
The shocked ionised gas then disperses subsonically, but to first order it is only free to move along field lines so it stretches into the filamentary structure seen in this figure.

\section{Conclusions}
Tests of implicit and explicit raytracing algorithms show that for simulations without rapidly moving ($v\gg10\,\mathrm{km}\,\mathrm{s}^{-1}$) or high temperature ($T\gg10^4$K) gas the implicit method is always more efficient on the uniform grid used here because of the large timesteps it allows.
By contrast, for a model with a fast stellar wind ($v=2000\,\mathrm{km}\,\mathrm{s}^{-1}$) the superior parallel scaling of the explicit algorithm makes it more efficient.

Our R-MHD photoionisation simulations show that both radiation-driven implosion and acceleration of clumps by the rocket effect tend to align the magnetic field with the radiation propagation direction in dense neutral gas (suggested by~\citealt{SugWatTamEA07} for the case of M\,16).
The effectiveness of this alignment is dependent on the initial field strength: with similar gas densities and pressures to those in M\,16, these simulations show that a magnetic field of $160\,\mu$G is sufficient to prevent any significant field reorientation.
By interpolating between the simulation results, an ambient field of $\vert\mathbf{B}\vert\lesssim50\,\mu$G is required to explain the observed field configuration in the M\,16 pillars \citep{SugWatTamEA07} if the pillars formed via the mechanism we are modelling.
If the field were stronger than this, additional signatures would also be seen in the ionised gas emission tracing the large-scale field orientation.
Such features appear to be present in H$\alpha$ images of the pillar near NGC\,6357 \citep{BohTapRotEA04} and in a pillar in the Carina nebula \citep[][Fig.\,1]{SmiBalWal10}.
These results show that magnetic field strengths in star-forming regions can in principle be significantly constrained by the morphology of structures which form at the borders of HII regions.

\acknowledgements JM's work has been part funded by the Irish Research Council for Science, Engineering and Technology, and by a grant from the Dublin Institute for Advanced Studies.
JM acknowledges the SFI/HEA Irish Centre for High-End Computing (ICHEC) for the provision of computational facilities and support.
JM acknowledges the John von Neumann Institute for Computing for a grant of computing time on the JUROPA supercomputer at Juelich Supercomputing Centre.

\bibliographystyle{asp2010}
\bibliography{jmackey}

\begin{thebibliography}{}
\expandafter\ifx\csname natexlab\endcsname\relax\def\natexlab#1{#1}\fi
\expandafter\ifx\csname url\endcsname\relax
  \def\url#1{\texttt{#1}}\fi
\expandafter\ifx\csname urlprefix\endcsname\relax\def\urlprefix{URL }\fi
\providecommand{\eprint}[2][]{\url{#2}}

\bibitem[{{Arthur} et~al.(2011){Arthur}, {Henney}, {Mellema}, {de Colle}, \&
  {V{\'a}zquez-Semadeni}}]{ArtHenMelEA11}
{Arthur}, S.~J., {Henney}, W.~J., {Mellema}, G., {de Colle}, F., \&
  {V{\'a}zquez-Semadeni}, E. 2011, \mnras, 414, 1747

\bibitem[{{Bohigas} et~al.(2004){Bohigas}, {Tapia}, {Roth}, \&
  {Ruiz}}]{BohTapRotEA04}
{Bohigas}, J., {Tapia}, M., {Roth}, M., \& {Ruiz}, M.~T. 2004, \aj, 127, 2826

\bibitem[{{Garc\'ia-Segura} \& {Franco}(1996)}]{GarSegFra96}
{Garc\'ia-Segura}, G., \& {Franco}, J. 1996, \apj, 469, 171

\bibitem[{{Gritschneder} et~al.(2010){Gritschneder}, {Burkert}, {Naab}, \&
  {Walch}}]{GriBurNaaEA10}
{Gritschneder}, M., {Burkert}, A., {Naab}, T., \& {Walch}, S. 2010, \apj, 723,
  971

\bibitem[{{Henney} et~al.(2009){Henney}, {Arthur}, {de Colle}, \&
  {Mellema}}]{HenArtDeCEA09}
{Henney}, W.~J., {Arthur}, S.~J., {de Colle}, F., \& {Mellema}, G. 2009,
  \mnras, 398, 157

\bibitem[{{Hester} et~al.(1996){Hester}, {Scowen}, {Sankrit}, {Lauer}, {Ajhar},
  {Baum}, {Code}, {Currie}, {Danielson}, {Ewald}, {Faber}, {Grillmair},
  {Groth}, {Holtzman}, {Hunter}, {Kristian}, {Light}, {Lynds}, {Monet},
  {O'Neil}, {Shaya}, {Seidelmann}, \& {Westphal}}]{HesScoSanEA96}
{Hester}, J.~J., {Scowen}, P.~A., {Sankrit}, R., {Lauer}, T.~R., {Ajhar},
  E.~A., {Baum}, W.~A., {Code}, A., {Currie}, D.~G., {Danielson}, G.~E.,
  {Ewald}, S.~P., {Faber}, S.~M., {Grillmair}, C.~J., {Groth}, E.~J.,
  {Holtzman}, J.~A., {Hunter}, D.~A., {Kristian}, J., {Light}, R.~M., {Lynds},
  C.~R., {Monet}, D.~G., {O'Neil}, E.~J., Jr., {Shaya}, E.~J., {Seidelmann},
  K.~P., \& {Westphal}, J.~A. 1996, \aj, 111, 2349

\bibitem[{{Lee} \& {Chen}(2007)}]{LeeChe07}
{Lee}, H., \& {Chen}, W.~P. 2007, \apj, 657, 884

\bibitem[{{Mackey}(2011)}]{Mac11}
{Mackey}, J. 2011, \aap, submitted

\bibitem[{{Mackey} \& {Lim}(2010)}]{MacLim10}
{Mackey}, J., \& {Lim}, A.~J. 2010, \mnras, 403, 714

\bibitem[{{Mackey} \& {Lim}(2011)}]{MacLim11b}
--- 2011, \mnras, 412, 2079

\bibitem[{{Mellema} et~al.(2006{\natexlab{a}}){Mellema}, {Arthur}, {Henney},
  {Iliev}, \& {Shapiro}}]{MelArtHenEA06}
{Mellema}, G., {Arthur}, S., {Henney}, W., {Iliev}, I., \& {Shapiro}, P.
  2006{\natexlab{a}}, \apj, 647, 397

\bibitem[{{Mellema} et~al.(2006{\natexlab{b}}){Mellema}, {Iliev}, {Alvarez}, \&
  {Shapiro}}]{MelIliAlvEA06}
{Mellema}, G., {Iliev}, I., {Alvarez}, M., \& {Shapiro}, P. 2006{\natexlab{b}},
  New Astronomy, 11, 374

\bibitem[{{Pound}(1998)}]{Pou98}
{Pound}, M.~W. 1998, \apjl, 493, L113+

\bibitem[{{Smith} et~al.(2010{\natexlab{a}}){Smith}, {Bally}, \&
  {Walborn}}]{SmiBalWal10}
{Smith}, N., {Bally}, J., \& {Walborn}, N.~R. 2010{\natexlab{a}}, \mnras, 405,
  1153

\bibitem[{{Smith} et~al.(2010{\natexlab{b}}){Smith}, {Povich}, {Whitney},
  {Churchwell}, {Babler}, {Meade}, {Bally}, {Gehrz}, {Robitaille}, \&
  {Stassun}}]{SmiPovWhiEA10}
{Smith}, N., {Povich}, M.~S., {Whitney}, B.~A., {Churchwell}, E., {Babler},
  B.~L., {Meade}, M.~R., {Bally}, J., {Gehrz}, R.~D., {Robitaille}, T.~P., \&
  {Stassun}, K.~G. 2010{\natexlab{b}}, \mnras, 406, 952

\bibitem[{{Sugitani} et~al.(2007){Sugitani}, {Watanabe}, {Tamura}, {Kandori},
  {Hough}, {Nishiyama}, {Nakajima}, {Kusakabe}, {Hashimoto}, {Nagayama},
  {Nagashima}, {Kato}, \& {Fukuda}}]{SugWatTamEA07}
{Sugitani}, K., {Watanabe}, M., {Tamura}, M., {Kandori}, R., {Hough}, J.~H.,
  {Nishiyama}, S., {Nakajima}, Y., {Kusakabe}, N., {Hashimoto}, J., {Nagayama},
  T., {Nagashima}, C., {Kato}, D., \& {Fukuda}, N. 2007, \pasj, 59, 507

\bibitem[{{Whalen} \& {Norman}(2008)}]{WhaNor08}
{Whalen}, D., \& {Norman}, M. 2008, \apj, 672, 287

\bibitem[{{Whalen} \& {Norman}(2006)}]{WhaNor06}
{Whalen}, D., \& {Norman}, M.~L. 2006, \apjs, 162, 281

\bibitem[{{Williams} et~al.(2001){Williams}, {Ward-Thompson}, \&
  {Whitworth}}]{WilWarWhi01}
{Williams}, R., {Ward-Thompson}, D., \& {Whitworth}, A. 2001, \mnras, 327, 788

\bibitem[{{Yorke}(1986)}]{Yor86}
{Yorke}, H.~W. 1986, \araa, 24, 49

\end{thebibliography}

\end{document}